# Analysis and Implementation of Distinct Steganographic Methods

Ünal TATAR and Tolga MATARACIOĞLU
TÜBİTAK UEKAE, Department of Information Systems Security
06700, Kavaklıdere, Ankara/TURKEY
{tatar,mataracioglu}@uekae.tubitak.gov.tr

*Abstract* - **In this paper, different steganographic methods have been analyzed and implementations of those techniques have been performed. Those methods include hiding in text, hiding in audio file, hiding in file system, and hiding in image files.**

*Index Terms* - **Steganography, steganographic methods, implementation, analysis.**

## I. INTRODUCTION

Steganography is the art of hiding data in data in an undetectable way. The word Steganography is derived from Greek and it means covered or hidden writing. Main concern of steganography is hiding existence of hidden message.

Steganography seems related with cryptography but it is different. Main concern in cryptography is hiding the content of the message but in steganography main concern is hiding the existence of hidden message.

Johannes Trithemius (1462-1516) has published a series of books named "Steganography: the art through which writing is hidden requiring recovery by the minds of men". The book I and Book II describe the methods to hide messages in writing. Book III is about secret astrology. Two researchers have discovered that Book III contains some hidden messages. One of those messages was "the quick brown fox jumps over the lazy dog".

Mary Queen of Scots has used the combination of steganography and cryptography so as to hide letters. She has used beer barrels which can easily pass in and out of her prison.

Another use of steganography is shaving a messenger's hair and writing the message on the head. After the messenger's hair grows up, the messenger can go to the intended receiver and the receiver also shaves his hair and retrieves the message. This method has been used by Histaiacus in 5th century BC.

In 480 BC, Demaratus warned Spartans about the risk of the invasion of the Xerxes. Heroclotus has decribed his method like:

"As the danger of discovery was great, there was only one way in which he could contrive to get the message through: this was by scraping the wax off a pair of wooden folding tables, writing on the wood underneath what Xerxes intended to do, and then covering the message over with the wax again."

In World War II, some steganographic methods have been used by Nazis. The name of the method was microdots. They were actually microfilm chips produced in high magnification. Those chips were the size of a period in a typewriter. However these periods contained very dense information, images, etc. The Nazis has also developed invisible inks as a method in steganography.

A Nazi spy has sent the information below:

"Apparently neutral's protest is thoroughly discounted and ignored. Isman hard hit. Blockade issue affects pretext for embargo on by-products, ejecting suets and vegetable oils."

When you pick the second letter in each word up, the following sentence occurs:

"Pershing sails from NY June I."

Cardano Grill, developed by Girolama Cardano, is another method in steganography. The grill is actually a piece of paper consists of holes. When this paper is put on the paper with information, the actual information can be retrieved from that paper by means of that grill.

The last ancient method described in this paper is using eyelids to blink words in Morse code. US Armed Forces prisoners in Vietnam have used this method by means of a five by five matrix. Each element of the matrix consists of a letter in the alphabet, and the Morse code of that letter. The following figure shows this five by five matrix:

|   | 1 | 2 | 3 | 4 | 5 |
|---|---|---|---|---|---|
| 1 | A . . | B . .. | C, K . …. | D . ….. | E . …... |
| 2 | F .. . | G .. .. | H .. … | I .. ….. | J .. …... |
| 3 | L …. . | M … .. | N … … | O … ….. | P … …... |
| 4 | Q …. . | R ….. .. | S ….. … | T ….. ….. | U ….. …... |
| 5 | V …... . | W …... .. | X …... … | Y …... ….. | Z …... …... |

Figure 1: The Morse code used by the US Armed Forces prisoners in Vietnam

## II. ANALYSIS AND IMPLEMENTATION

There are different types of steganography. Main difference between techniques is using different carriers. In application, a text message, an image file, an executable program file or an audio file can be used as a carrier. In this part of paper, we will introduce steganography tecniques with different types of carriers.

*A. Hiding in Text*

In this part, text based steganography, as another method in steganography, will be explained briefly.

Figure 2 illustrates the block diagram of text based steganography.

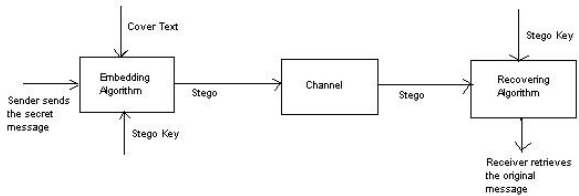

Figure 2: Block diagram of text based steganography

Firstly, sender sends the secret message by using the embedding algorithm, stego key and the cover text which will conceal the original message and the stego text is formed. Then this text is transmitted over the channel and comes to the receiver. The receiver uses the recovering algorithm and the stego key so as to retrieve the original message from the stego text.

Text based steganography can be classified into three main categories: Format-based, Random and statistical generation and linguistic method.

Generally, format-based method modifies existing text so as to hide the steganographic text. Random and statistical generation method is generating cover text according to the statistical properties. This method is based on character sequences and words sequences. Linguistic method specifically considers the linguistic properties of generated and modified text, frequently uses linguistic structure as a place for hidden messages.

One example of text steganography is to turn a short message into a spam mail content, send the mail to the receiver and retrieve the original message. We have given the following message as the input to the program in:

*"This is a secret message!"*

The program has turned this input into a spam mail content as below:

*"Dear Friend; Thank-you for your interest in our publication . If you no longer wish to receive our publications simply reply with a Subject: of "REMOVE" and you will immediately be removed from our club ! This mail is being sent in compliance with Senate bill 1816 ; Title 3 ; Section 304 . This is not multi-level marketing . Why work for somebody else when you can become rich within 45 days . Have you ever noticed more people than ever are surfing the web & people love convenience ! Well, now is your chance to capitalize on this . We will help you SELL MORE and use credit cards on your website . You are guaranteed to succeed because we take all the risk ! But don't believe us ! Ms Ames of North Carolina tried us and says "I was skeptical but it worked for me" ! We are licensed to operate in all states . We IMPLORE you - act now . Sign up a friend and you'll get a discount of 80% ! Cheers . Dear Cybercitizen , You made the right decision when you signed up for our mailing list ! This is a one time mailing there is no need to request removal if you won't want any more . This mail is being sent in compliance with Senate bill 1621 , Title 2 ; Section 301 ! THIS IS NOT A GET RICH SCHEME . Why work for somebody else when you can become rich within 74 weeks . Have you ever noticed people love convenience & nobody is getting any younger . Well, now is your chance to capitalize on this . WE will help YOU decrease perceived waiting time by 130% plus increase customer response by 150% ! You can begin at absolutely no cost to you . But don't believe us ! Ms Anderson who resides in Alabama tried us and says "Now I'm rich many more things are possible" ! This offer is 100% legal ! So make yourself rich now by ordering immediately ! Sign up a friend and you'll get a discount of 60% . Best regards !"*

*B. Hiding in Audio File*

Audio steganography and spectral estimation methods will be explained briefly as the subject of this part of the paper.

In a computer-based audio steganography system, secret messages are embedded in digital sound. The secret message is embedded by slightly altering the binary sequence of a sound file. Existing audio steganography software can embed messages in WAV, AU, and even MP3 sound files. Embedding secret messages in digital sound is usually a more difficult process than embedding messages in other media, such as digital images.

Another aspect of audio steganography that makes it so attractive is its ability to combine with existing cryptography technologies. Users no longer have to rely on one method alone. Not only can information be encrypted, it can be hidden altogether.

In order to estimate the power spectra of the signals in Additive White Gaussian Noise, there exists some estimation methods. Some of those are The Periodogram Method, The Blackman and Tuckey Method, Capon's Method, Yule-Walker Method, and Modified Covariance Method.

To compare, it can be said that the Blackman-Tuckey estimates perform better than the corresponding Periodogram estimates. This means that the variance in Blackman-Tuckey estimate curves is smaller than that of the Periodogram estimate curves.

In our implementation, a real data is concerned to analyse the practical situation. This data is a sound data and it is transmitted to the MATLAB by means of the Standard MATLAB function given below:

*signal=wavread('sample.wav');*

After performing this line, the original sound is transmitted to a vector. To give sound to the data vector, the standard MATLAB function given below is necessary:

*sound(signal);*

The waveform of the sound used for this paper is given in Figure 3:

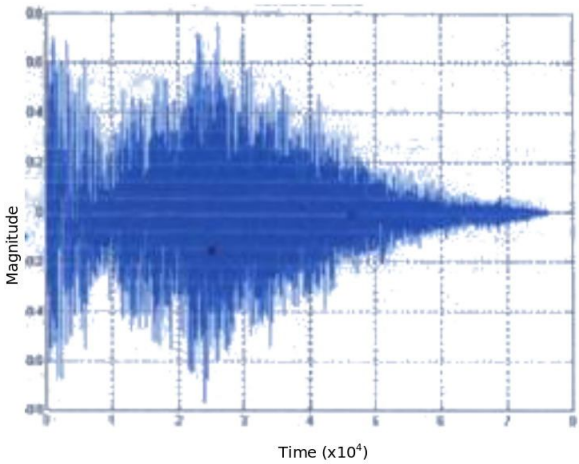

Figure 3: The waveform of the sound

And let our signal be:

$$x(n) = \cos(0.4\pi n) + sound(n)$$

It can be observed from Figure 4 that, all the methods give worse performance since the noise part of the signal (here, the sound) is correlated and its shape is not Gaussian distributed. However the Modified Covariance Method estimates the frequency.

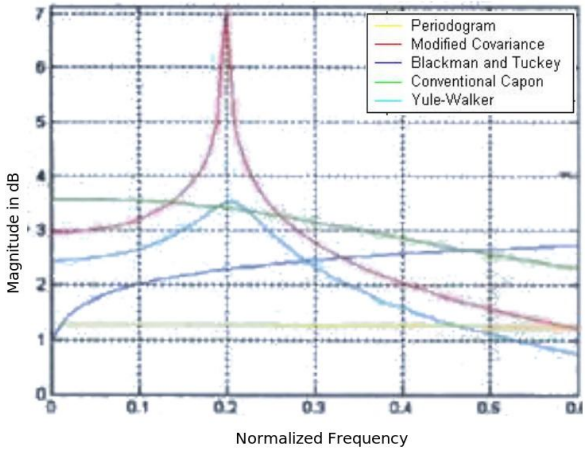

Figure 4: Estimation of the frequencies for the specified signal

By means of this simulation, we can say that we can add our information signal to a real data and send it to the receiver through a channel and the receiver takes the data and applies the best estimator to pull out the information from the real data.

In conclusion, as more emphasis is placed on the areas of copyright protection, privacy protection, and surveillance, we believe that steganography will continue to grow in importance as a protection mechanism. Audio steganography in particular addresses key issues brought about by the MP3 format, P2P software, and the need for a secure broadcasting scheme that can maintain the secrecy of the transmitted information, even when passing through insecure channels.

### C. Hiding in File System

In this section of the paper, file system steganography will be explained. Slack spaces, alternate data streams and merging files in other files are the subsections of this section.

A file system uses fixed size of space to store files in chunks of data which is known as blocks in Unix/Linux systems, and clusters in Windows systems. The smallest process unit is called sector and one sector consists of 512 bytes. Clusters occur when multiple sectors conjugate. Also the cluster size varies with the type of the operating system. Files are recorded in clusters.

There exist 2 different cases, one is the size of the file and the cluster are the same. This is the ideal case and the probability of the occurance of this case is extremely low. The second case is the size of the file is larger or smaller than the cluster size. In this case, it is certain that the file will not fit into the cluster. There will be overflows to other clusters. The space between the end of the file and the end of the cluster is called the slack space and the information hidden in this space is called slack space data . For instance, if a file is 200 bytes long, the file has to allocate a 2048-byte data unit, and the remained 1848 bytes will be slack space considering an NTFS file system with a 2048-byte cluster. Figure 5 illustrates the slack space.

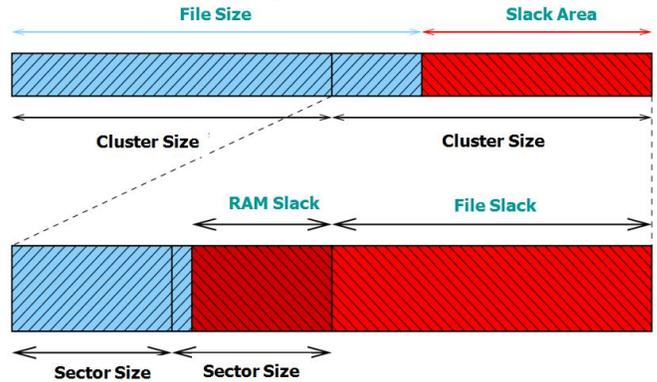

Figure 5: Illustration of slack space

Slack area or slack space consists of two types of slacks: RAM slack and File slack. Remember that clusters contain multiple sectors. RAM slack is defined as the space between the end of the file and the end of the sector. File slack is defined as the space between the beginning of the next sector and the end of the cluster.

When you press delete button on a file in a computer, you actually do not delete the whole file, only the pointer which indicates the beginning of the file is deleted. If this pointer is deleted, then the computer may write another information in that data unit. Slack space is not emptied in most computers, so this space contains information or data from previous files or from memory. Those data can be used by baleful people. Also this space may be used in steganographic methods. However, it is probable that the operating system may overwrite on those slack areas.

Another method of hiding information is file systems is to use hidden partitions. If the user knows the file name and password, then he/she can access to the file, otherwise the file seems not to exist in the system.

The aim of the Metasploit Anti-Forensics Project, developed by Vinnie Liu, is to develop tools so as to remove forensic evidence from computers. One of those tools is slacker.exe. by means of this executable, one can hide messages by using slack areas in NTFS file systems. Slacker.exe can be reached from and the usage is given in Figure 6.

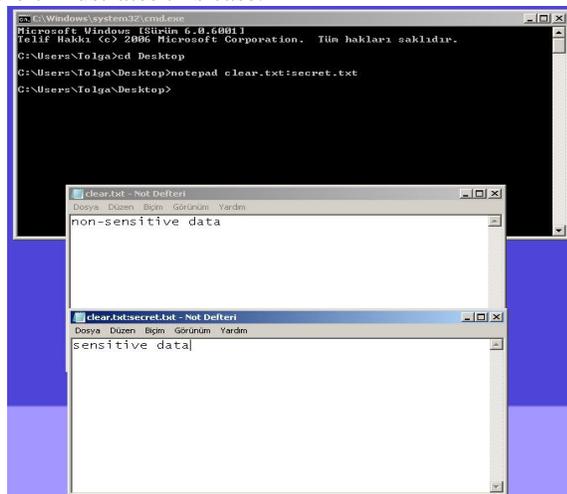
Figure 6: Use of slacker.exe

We have already used notepad.exe in Windows operating systems. Now let us examine how to hide information by using this exe. Our aim is to hide information stored in secret.txt by using clear.txt. We have non-important information in clear.txt and sensitive data in secret.txt. so as to create secret.txt, go to the folder where clear.txt exists in the command prompt and write the following command:
notepad clear.txt:secret.txt

Thereby, you cand create a txt file and write your sensitive data down in it, however you can not see this file in the computer. In order to see the content of this file, go to the command prompt and write the command above again. Figure 7 illustrates this case.

Figure 7: clear.txt and secret.txt files

This property comes with NTFS file systems and is called alternate data streams (ADS). If you want to copy this file into a system using FAT32 (most of the USB memory cards use this type of file system), you will get the warning window in Figure 8.

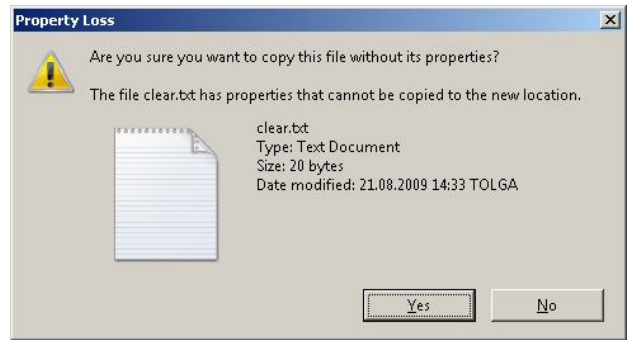
Figure 8: Copying ADS from NTFS to FAT32

Now let us do the same exercise by using two different executables. We will use winmine.exe (minesweeper) and calc.exe (calculator). You may consider that winmine.exe is an innocent program, however calc.exe is a malicious program like a virus will merge calc.exe into winmine.exe. Figure 9 gives us the commands so as to do this:

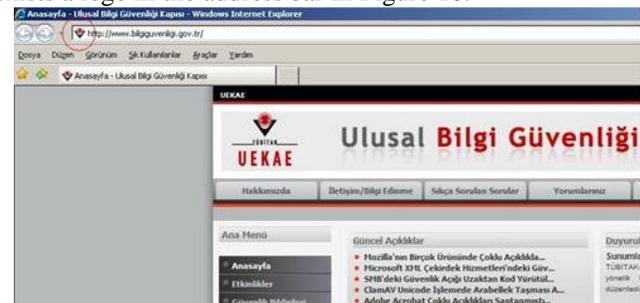
Figure 9: Merging calc.exe into winmine.exe

Finally, let us explain how the files with alternate data streams can be detected and removed. One of the famous tools for this purpose is Microsoft's streams.exe. Notice that there exists a logo in the address bar in Figure 10:

Figure 10: The logo

After adding this site to the Favorites in Internet Explorer, run the first command in Figure 11, so you can detect ADS's in the relevant directory.

Figure 11: Detection of ADS's

One can see the line about the logo in the last line in Figure 7 and conclude that this logo has been merged in bilgiguvenligi.gov.tr URL. If you open that file in notepad, you may see the merged information seen in Figure 12.

Analysis and Implementation of Distinct Steganographic Methods 5

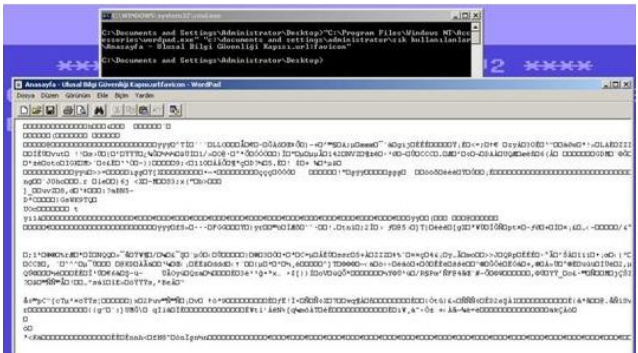

Figure 12: Logo format seen in notepad

So as to remove those ADS's, you can run the same command by using –d parameter.

### D. Hiding in Image Files

Hiding data into the image files is another type of steganography. This method is the most common steganographic technique. There are several methods for hiding data into images. Most common methods for image steganography are least significant bit method and masking and filtering

In this project paper we will deal with only least significant bit method.

#### 1) Image Files

Each image file in digital form is composed of pixels. Every pixel is similar to a very small point in picture. A pixel is either 24-bit or 8-bit. A common image size is 640X480 and such a file is 300 000 pixels.

Each pixel represents a color and each color is composed of red, green and blue. Different amount of these colors defines the value of pixel in 1 and 0s. For a 24-bit image, 8-bit is used for red, green and blue, total is 24 bits. This means that each pixel can have 16,777,216 ($2^{24}$) different colors.

Image files, like jpg files, can be compressed. Compressed files have two forms of compression: lossless and lossy compression. In lossless compression, you can retrieve the original file but in lossy compression you can not retrieve original file but only compressed form.

#### 2) Least Significant Bit (LSB) Method

We have two files in hiding information in an image file: Cover image which is carrier of the hidden message and the other file is hidden message. After inserting the hidden message into cover image we have the stego image file.

In image steganography most common technique is least significant bit insertion technique. For instance, we want to hide letter 'A' in three pixels data. 'A' letter has 8 bits data. 'A' is equal to ( 1 0 0 0 0 0 1 1 ) in binary format.

Each pixel has 24 bit information. Assume following is the value of three pixels.

( 00100111 11101001 11001000 )
( 00100111 11001000 11101001 )
( 11001000 00100111 11101001 )

To hide data, we change last bits of each byte (8 bits), that means least significant bits of Red, Green and Blue colors of each pixel.

After hiding data, data in pixels is shown below.

( 0010011*1* 1110100*0* 1100100**0** )

( 0010011*0* 1100100**0** 1110100*0* )

( 1100100**0** 0010011*1* 1110100**1** )

In this example, only italic and bold digits changed, red ones are unchanged. Since data and cover image are composed of 1's and 0's probability of change of a bit cover image is about ½.

If we use last two bits of RGB values of each pixel as in Figure 13, we can hide two times larger data in the image file and change in image is still undetectable by human eyes.

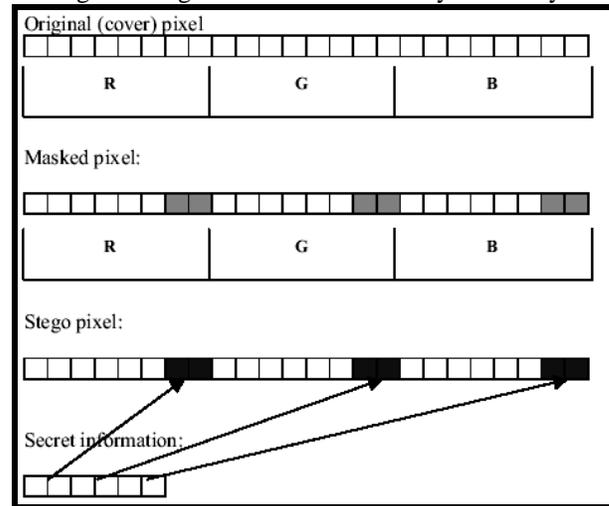

Figure 13: Hiding in last two bits.

#### 3) An LSB Application

For this paper we implemented an LSB application. The program uses only 24-bit bmp image as cover media. In Figure 14, screenshot of the program is seen.

To hide a file:

1. Load Image: Click Load Image button on the left pane to select cover image file.

2. Load: Click load button to select guest data (hidden file)

3. Passphrase: Write a password to encrypt (DES algorithm) hidden data before hiding. This is optional.

4. Used Bits: Select Used Bits ruler to select how many bits used.

5. Hide Data: Click Hide Data button to hide hidden data in cover image

6. Save Image: Click Save Image button to save the modified image file.

To extract hidden data:
1. Load Image: Click Load Image button on the right pane to select image that contains hidden data.

2. Passphrase: Write the password to decrypt (DES algorithm) hidden data.

3. Used Bits: Select Used Bits ruler to select how many bits used.

4. Extract Data: Click Extract Data button to unhide the data.

5. Save: Click Save button to save the hidden data.

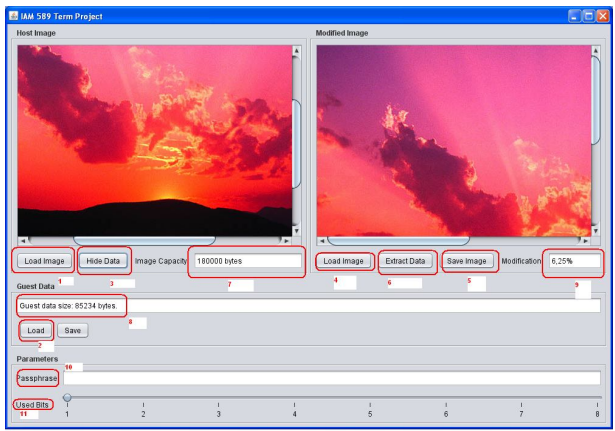

Figure 14: Screenshot of the LSB Steganography Program

When we used more than one bit deformation of the image increases. In the Figure 15 we hide a file in the same image file and we used last bits from 1 to 8.

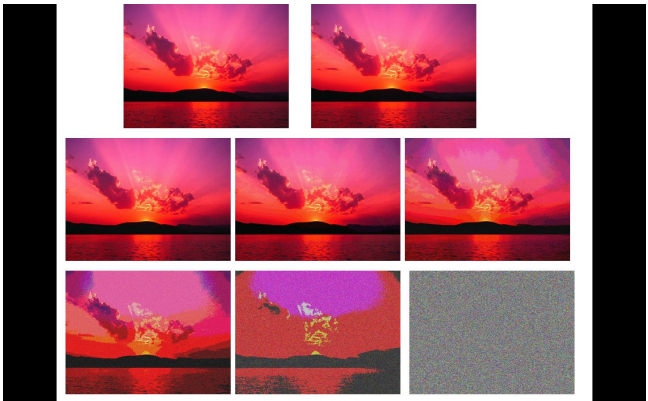

Figure 15: Corruption in cover image files

### III. CONCLUSION

This paper aims to analyze distinct steganographic methods including hiding in text, hiding in audio file, hiding in file system, and hiding in image files. Further, implementations of those techniques have been performed.

### REFERENCES

[1] A. Ker: "Improved Detection of LSB Steganography in Grayscale Images", in J. Fridrich (ed.): Information Hiding. 6th International Workshop. Lecture Notes in Computer Science, vol. 3200, Springer-Verlag New York, pp. 97–115, 2005.

[2] Avcibas, I.; Memon, N.; Sankur, B. "Image steganalysis with binary similarity measures"

[3] B. Carrier, "File System Forensic Analysis", Addison-Wesley Publishing, 2005.

[4] "Classical Steganography, Cardano Grille" URL: http://library.thinkquest.org/27993/crypto/steg/classic1.shtml

[5] Counterintelligence News and Developments, Volume 2, June 1998 "Hidden in Plain Sight-Steganography" URL:www.nacic.gov/pubs/news/1998/jun98.htm

[6] S. Dumitrescu, X. Wu, and Z. Wang: "Detection of LSB Steganography via Sample Pair Analysis", in: Petitcolas, F.A.P. (ed.): Information Hiding. 5th International Workshop. Lecture Notes in Computer Science, vol. 2578, Springer-Verlag New York, pp. 355–372, 2000.

[7] Fabien A. P. Petitcolas, Ross J. Anderson and Markus G. Kuhn. "Information Hiding – A Survey", Proceedings of the IEEE, special issue on protection of multimedia content, July 1999, pp. 1062 – 1078.

[8] H. Farid and L. Siwei: "Detecting Hidden Messages Using Higher-Order Statistics and Support Vector Machines", in: F.A.P. Petitcolas (ed.): Information Hiding. 5th International Workshop. Lecture Notes in Computer Science, vol. 2578, Springer-Verlag New York, pp.340–354, 2002

[9] I. Avcibas, N. Memon, and B. sankur, \Steganalysis using image quality metrics." *Security and Watermarking of Multimedia Contents, San Jose, Ca.*, Feruary 2001.

[10] J. Fridrich, M. Goljan, D. Hogea, and D. Soukal: "Quantitative Steganalysis: Estimating Secret Message Length," ACM Multimedia Systems Journal, Special Issue on Multimedia Security, vol. **9**(3), pp. 288–302, 2003.

[11] J. Zollner, H. Federrath, H. Klimant, A. P¯tzman, R. Piotraschke, A. Westfeld, G. Wicke, and G. Wolf, \Modeling the security of steganographic systems," *2nd Information Hiding Workshop*, pp. 345{355, April 1998.

[12] K. Bennett, "Linguistic Steganography: Survey, Analysis, and Robustness Concerns for Hiding Information in Text", Purdue University, CERIAS Tech. Report, 2004.

[13] Kay, S. M., "Modern Spectral Estimation Theory and Application", Prentice Hall, Jan. 1988.

[14] Kolata, Gina, "A Mystery Unraveled, Twice" URL: cryptome.unicast.org/cryptome022401/tri.crack.htm

[15] M. Russinovich. Streams v1.56. URL: http://technet.microsoft.com/en-us/sysinternals/bb897440.aspx

[16] Miroslav Goljan∗ , Jessica Fridrich, and Taras Holotyak "New Blind Steganalysis and its Implications "

[16] N.F. Johnson, Z. Duric, S. Jajodia, "Information Hiding: Steganography and Watermarking – Attacks and Countermeasures", Kluwer Academic Publishers, 2001.

[17] R. Anderson, R. Needham, A. Shamir, "The Steganographic File System", Lecture Notes in Computer Science, vol. 1525, Sprinter-Verlag, 1998.

[18] Spam Mimic. URL: http://www.spammimic.com/index.shtml

[19] Singh, Simon, "The Cipher of Mary Queen of Scots" URL: www.arch.columbia.edu/DDL/cad/A4513/S2001/r9/

[20] S. Lyu and H. Farid "Steganalysis Using Higher-Order Image Statistics" IEEE Transactions on Information Forensics and Security, 1(1):111-119, 2006

[21] Taras Holotyak1, Jessica Fridrich1, Sviatoslav Voloshynovskiy2 "Blind Statistical Steganalysis of Additive Steganography Using Wavelet Higher Order Statistics"

[22] T. Mataracıoğlu, "Uygulamalarla Steganografi", National Information Security Portal. URL: http://www.bilgiguvenligi.gov.tr/teknik-yazilar-kategorisi/uygulamalarla-steganografi.html?Itemid=6

[23] T. Mataracıoğlu, Ü. Tatar, "Spectral Estimation Methods: Comparison and Performance Analysis on a Steganalysis Application", 2nd Information Security & Cryptology Conference with International Participation, Ankara, Dec. 2007.

[24] Trithemius, Johannes, "Steganographia:hoe est ars per occultam scripturam animi sui voluntatem absentibus aperiendi certa" URL: www.esotericarchives.com/tritheim/stegano.htm

[25] "The Science of Secrecy, Steganography" URL: www.channel4.com/plus/secrecy/page1b.html

[26] Van Trees, H. L., "Detection, Estimation and Modulation Theory Part-I", John Wiley, Jan. 1968.

[27] V. Liu, Metasploit Anti-Forensics Project. URL: http://metasploit.org/research/projects/antiforensics